\documentstyle[12pt]{article}

\begin{document}

\begin{flushright}
MRI-PHY/P971029 \\
hep-th/9710035
\end{flushright} 

\vspace{2mm}

\vspace{2ex}

\begin{center}
{\large \bf  Size of Black Holes through Polymer Scaling} \\ 

\vspace{6mm}
{\large  S. Kalyana Rama}
\vspace{3mm}

Mehta Research Institute, Chhatnag Road, Jhusi 

Allahabad 221 506, India. 

\vspace{1ex}
email: krama@mri.ernet.in \\ 
\end{center}

\vspace{4mm}

\begin{quote}
ABSTRACT. 
We imagine the strings on the strectched horizon of any 
$d$ space-time dimensional black hole to be bits of polymer. 
Then, proposing an interaction between these bits we obtain 
the size of the configuration, and thus of the black hole, 
using the scaling laws. The transition from a typical black 
hole state to a typical string state has a simple 
explanation, which also holds for the extremal black holes. 
\end{quote}

\newpage

Recently, tremendous progress has been made in understanding 
the entropy of extremal and nonextremal black holes in string 
theory. See \cite{review} for a recent review. However, 
understanding of the properties of the Schwarzschild black holes 
still remains incomplete. Recently proposed (M)atrix theory 
\cite{matrix} appears to be poised to provide an answer to 
this problem \cite{banks}. 

In a series of insightful papers, Susskind had shown that 
the transverse size of the strings on the stretched horizon 
(sometimes referred to simply as horizon in the following) 
increases and covers the entire horizon, and argued that 
the black hole entropy arises from such configuration of 
strings \cite{sh}. Linking these two phases is 
the correpondence principle \cite{hp} which states that when 
the horizon size becomes of the order of the string scale 
a typical black hole state smoothly transforms into 
a typical string state. This transition has been analysed 
further in \cite{hp2}. 

The (stretched) horizon can be defined as the location from 
where the red shift factor is of the order of inverse size of 
the black hole. It is shown that the transverse spreading of 
the strings on the horizon can be described as a branching 
process diffusion of string bits, the seperation between 
the adjacent bits being of the order of the string scale 
$a \equiv \sqrt{\alpha'}$ \cite{peet}. 

In this note, we imagine the strings on the strectched horizon 
of any $d$ space-time dimensional black hole to be bits of 
polymer. We assume that $d \ge 4$. 
Then, proposing an interaction between these bits 
assumed to be valid at any value of string coupling $g$, 
we obtain the size of the configuration using the scaling 
laws \cite{degennes,cardy}. \footnote{For a different use of 
the idea of strings as polymers, see \cite{thorn}. The scaling 
concepts have also been used recently in \cite{hp2}.}
This size can be interpreted as the size of the horizon, that 
is the size of the black hole, and has the correct dependence 
on the mass. The entropy of the configuration has also the 
same dependence on the mass as for a black hole. It is also 
possible to see that the size of the extremal black hole is 
naturally of the string size \cite{hawking}. However, as is 
common in the application of scaling laws, the exact 
coefficients can not be ontained, except through an adhoc 
input of coefficients. 

It is known that at vanishingly small values of string coupling 
$g$, the length $R_0$ of the string scales as 
\begin{equation}\label{r0}
R_0^2 = a^3 M 
\end{equation}
where $a = \sqrt{\alpha'}$ is the string scale, and $M$ is 
the string mass \cite{peet,turok}. When $g$ becomes strong, 
the strings interaction becomes strong. It is then natural to 
assume that the strings split into a number of bits ($ = n$), 
with an average length $= a$, and each bit behaving 
independently. This collection can be thought of as a collection 
of polymers. It is then reasonable to apply to this system mean 
field theory and scaling law arguements, such as the ones used 
in polymer physics \cite{degennes,cardy}. 

The number of bits $n$ can be determined by requiring that when 
$g$ is small, the total mean square length of these $n$ bits 
be equal to $<R_0^2>$ given in (\ref{r0}). Since the mean square 
length of each bit $= a^2$, this gives 
\begin{equation}\label{n}
n a^2 = a^3 M \; , \; \; \; \; {\rm i.e.} \; \; \; \; 
n = a M \; . 
\end{equation} 
It follows that the mass of each bit $= a^{- 1}$. 

Now one can follow Flory's approach \cite{degennes} and 
write the free energy $F$ for the above system in terms 
of the size $R$ of the bits, including interactions. Minimising 
$F$ then gives a relation for $R$. Note that, upto constant 
coefficients, 
\begin{equation}\label{mts}
F_{{\rm min}} \simeq M \simeq T S \; , 
\end{equation}
where $T$ is the temperature and $S$ is the entropy. 
The free energy for $n$ bits 
can be written, in the absence of interactions, as 
\cite{degennes,cardy,guy} 
\begin{equation}\label{f0}
F_0 = n T \left( \frac{R^2}{a^2} + \frac{a^2}{R^2} \right) 
\end{equation}
where $T$ is the temperature. Here and in the following 
the expressions are correct upto some constant coefficients, 
which suffices to obtain qualitative features. $F_0$ is 
minimum and $= n T$ at $R = a$. Requiring that $F_0 ({\rm min}) 
\simeq M$ and using (\ref{n}) gives 
\begin{equation}\label{t0}
T = a^{- 1} \simeq T_H \; , 
\end{equation}
where $T_H$ is the Hagedorn tempearture. From (\ref{mts}) 
it then follows that the entropy S is given, upto a constant 
coefficient, by 
\begin{equation}\label{s0}
S = a M \; . 
\end{equation}
Note that the second term in (\ref{f0}) can be neglected when 
$R >> a$, as will be the case in the presence of interactions. 

When the interactions are included, the free energy 
can be written in the form 
\begin{equation}\label{f1}
F = n T \frac{R^2}{a^2} + n T \tilde{F}_I \; , 
\end{equation}
where we have omitted the second term of (\ref{f0}). 
This free energy describes the dynamics of the string bits 
on the stretched horizon. For an asymptotic observer, however, 
the free energy is red shifted by a factor $\frac{a}{R}$ as 
follows from the definition of stretched horizon \cite{sh}. 
Thus, the asymptotic free energy is 
\begin{equation}\label{finf}
F_{\infty} = \frac{a}{R} \left( 
n T \frac{R^2}{a^2} + n T \tilde{F}_I \right) \; . 
\end{equation}
But $F_{\infty}$(min), upto a coefficient, must be equal to 
the mass $M$ of the system. When $F_{\infty}$ is minimised, 
both the terms in (\ref{finf}) will turn out to be of the same 
form. Therefore, requiring $F_{\infty}$(min) $\propto M$, 
and using (\ref{n}), then determines $T$ upto a coefficient: 
\begin{equation}\label{t}
T = R^{- 1} \; . 
\end{equation}

Thus, the free energy of the string bits at the strectched 
horizon can be written as 
\begin{equation}\label{fi'}
F = n \frac{R}{a^2} + n F_I \; , 
\end{equation}
where $F_I$ is the effective interaction felt by 
one string bit due to the mean field of the others 
at the stretched horizon. 

We now propose that on the strectched horizon 
(i) the string bits effectively live in 
a $(d - 1)$ dimensional space-time; (ii) the effective 
interaction between the bits is actually repulsive; and 
(iii) its form is that of (Newtonian) gravitational potential 
in $(d - 1)$ dimensional space time. This implies that, 
in the mean field theory approach, each bit contributes 
to free energy additively an amount equal to
\begin{eqnarray}
& & a^{- 1} G_{d - 1} M R^{4 - d} \; , \; \; \; 
(d \ne 4) \nonumber \\ 
& & - a^{- 1} G_3 M \ln R \; , \; \; \; (d = 4) \; , 
\label{fi1}
\end{eqnarray} 
where $G_{d - 1}$ is the Newton's constant in $(d - 1)$ 
dimensional space time, $a^{- 1}$ is the mass of one bit, 
$R$ is the size of the system, and $M$ is the mass 
of the remaining bits. For $n$ bits, the contribution is 
$n$ times the above expression. 

In string theory, $G_{d - 1} = g^2 a^{d - 3}$ where 
$g$ is the string coupling constant. Therefore, 
including the interaction energy of $n$ bits, and using 
(\ref{n}), the free energy (\ref{f1}) becomes 
\begin{eqnarray}
F & = & M \frac{R}{a} + g^2 M^2 \frac{a^{d - 3}}{R^{d - 4}} 
\; , \; \; \; (d \ne 4) \nonumber \\ 
& = & M \frac{R}{a} - g^2 M^2 a \ln R 
\; , \; \; \; (d = 4) \; . \label{fi}
\end{eqnarray}
Note that the expression is reliable only upto constant 
coefficients. 

Another equivalent way to treat the $d = 4$ case in (\ref{fi1}) 
and (\ref{fi}) is to put $d = 4 + \epsilon$ in the 
general expression, and (un)renormalize the coupling constant 
$g^2$ to $g_0^2$ as follows: 
\begin{equation}\label{ren}
\frac{1}{g^2} = \frac{1}{g^2} \left( \frac{1}{\epsilon} 
+ \cdots \right) \; ,  
\end{equation} 
where $\cdots$ are subleading terms. This is similar to 
the renormalization of $G$ in the work of Susskind and Uglum 
in \cite{sh}. The limit $\epsilon \to 0$ is taken in the end. 

Equation (\ref{fi}) gives the free energy of the system of 
string bits living on the stretched horizon. The size of 
the system, and hence of the horizon, is now easily obtained 
by minimising $F$ with respect to $R$. The result is 
\begin{equation}\label{r}
R \propto M^{\frac{1}{d - 3}} \; , 
\end{equation}
which is the actual case for the black holes. Using now 
(\ref{mts}), (\ref{finf}), (\ref{t}), and (\ref{r}), 
it now follows that the entropy $S$ is given by 
\begin{equation}\label{s} 
S \propto R^{d - 2} \; \; \propto M^{\frac{d - 2}{d - 3}} \; . 
\end{equation}
Equation (\ref{ren}) is also used when $d = 4$. 

The scaling of $R$ as in (\ref{r}) can also be derived from 
Edwards Hamiltonian (see chapter 9 of \cite{cardy}). It 
follows from (\ref{fi}) that 
\begin{equation}\label{edf}
a^{- 1} \frac{F}{T} = M \frac{R^2}{a^2} 
+ g^2 M^2 \frac{a^{d - 4}}{R^{d - 5}} \; . 
\end{equation}
(We treat the $d = 4$ case as $d = \lim_{\epsilon \to 0} 
(4 + \epsilon)$ and use the equation (\ref{ren}) for 
the coupling constant.)  This suggests that the corresponding 
Edwards Hamiltonian is given by 
\begin{equation}\label{edh}
{\cal H} = M \int d t \left( \frac{d R}{d t} \right)^2 
+ g^2 M^2 a^{d - 4} \int \int d t_1 d t_2 
\vert R_1 - R_2 \vert^{5 - d} \; , 
\end{equation}
where $t$ is a parameter \cite{cardy}. 
Now we follow the scaling argument given by Cardy \cite{cardy}. 
In our case note that a scaling $T \sim a$ is necessary to make 
the first term independent of $a$. This reflects that the length 
of the $n$ bits, in the absence of interaction, is given by 
$R \sim a \left( \frac{t}{a} \right)$. Setting $t = n$, one gets 
that the length of each bit is $\sim a$, see (\ref{f0}). 

To obtain the scaling of $R$ with respect to $t$, assume that 
$R$ rescales as $b^{- x} R$ under a rescaling $a \to b a$. 
Thus $R (t)$, which must have the form $a f(\frac{t}{a})$ 
by dimensional analysis, scales with $t$ as $t^{1 + x}$. 
Now the two terms in (\ref{edh}) scale as $b^{- 2 x}$ 
and $b^{d - 4 + (d - 5) x}$ respectively. Equating 
the exponents gives $x = \frac{4 - d}{d - 3} = - 1 
+ \frac{1}{d - 3}$. Setting $t = n$, it follows that 
\begin{equation}\label{scale}
R \sim n^{\frac{1}{d - 3}} \; ,
\end{equation}
which is same as (\ref{r}) as follows from {\ref{n}). 

In this approach, there is a simple explanation for 
the transition from a typical black hole state 
to a typical string state as the coupling constant $g$ 
is decreased. As $g$ decreases, the interaction term 
$F_I$ in (\ref{fi}) decreases and in the limit of vanishing 
$g$ it becomes negelgible and can be dropped. The free 
energy then is the same as in (\ref{f0}), from which 
it follows that the size of the string bit $= a$ and that 
of $n$ bits is given by (\ref{r0}). The entropy is then 
given by (\ref{s0}). 

Note that the above result is true whenever the interaction 
term is absent. This suggests a simple explanation for 
the size and the entropy of an extremal black hole. 
It is known that the extremal black holes exert no force 
on each other since, in the extremal case, the gravitational 
force is exactly cancelled by the coulomb force (and scalar 
forces, if any). This can be seen as a consequence of 
the saturation of BPS limit, or of extended (super)symmetries. 
It is therefore natural that the string bits do not interact 
and, hence, the interaction term in (\ref{fi}) should be 
completely absent. This complete absence of interaction 
may perhaps be ensured by certain symmetries. The result is 
then the same as in the above case, which is consitent with 
the recent results \cite{hawking,ashoke,vafa}. 

However, it is not clear how to treat the charged (non extremal) 
black holes. A simple way would be to replace $g^2 M^2$ in 
(\ref{fi}) by $g^2 M^2 - Q^2$ where $Q$ is the charge of 
the black hole. But this does not lead to the correct 
scaling of the size $R$ with the mass $M$ and the charge $Q$. 
The scaling obtained by the above simple replacement may, 
perhaps, be correct only in the limit $Q \to 0$. 

Although simple, the present proposal has many shortcomings. 
Some of them are: (i) it is not clear if/whether/how the strings 
split into bits as the coupling $g$ is increased; 
(ii) why the gravitational interaction on the stretched horizon 
is repulsive; (iii) the meaning of redshift factor within 
the present framework; (iv) the meaning of temperature 
$T$ in (\ref{t}) derived using this factor. Most important, 
and perhaps a fatal, shortcoming is (v) the coefficients 
needed for various terms in the free energy in (\ref{fi}). 

On the other hand, the results of \cite{sh,peet,turok} almost 
naturally leads to imagining strings as bits and treating them 
as polymers. \footnote{One may perhaps treat them a gas of 
string bits, but then the first term in (\ref{fi}) does not 
appear to have a simple explanation.} This approach leads to 
a derivation of the size of the black holes. It may also 
be taken to give correct the functional forms of temperature 
and entropy. The transition from a typical black hole state 
to a typical string state has a simple explanation, which also 
holds for the extremal black holes. Lastly, since the 
Schwarzschild black holes have remained difficult to understand 
it may be worth to examine the present proposal in more detail.

\end{document}